\documentclass[12pt]{article}

\usepackage{graphicx}
\usepackage{amssymb}

\begin{document}

\title{$dd\to {^3}He n$ reaction at intermediate energies
\thanks{supported by the Russian Foundation for Basic Research
under grant  No.  10-02-00087a}
}


\begin{center}
N. B. Ladygina

 Joint Institute for Nuclear Research, LHEP, 141980, Dubna, Russia \\

{\it E-mail: nladygina@jinr.ru}
\end{center}
%


\begin{abstract}
The  $dd\to ^3He n$ reaction is considered at the energies between
200 MeV and 520 MeV. The Alt-Grassberger-Sandhas equations are
iterated up to the lowest order terms over the nucleon-nucleon t-matrix.
The parameterized ${^3He}$ wave function including five components
is used. The angular dependence of the differential cross section 
and energy dependence of tensor analyzing power $T_{20}$
at the zero scattering angle are presented in comparison with
 the experimental data.
\end{abstract}

\section{Introduction}
\label{intro}

During several  decades hadronic  reactions with  helium and tritium were extensively investigated
 at the energies of a few hundred MeV. A number of experiments to study 
  nucleon knockout from polarized ${^3He}$ were performed
at TRIUMF. As a result, differential cross sections  were measured at the energies between 220 and 590 MeV \cite{epst},\cite{brash},\cite{kit}. 
 Moreover,
the polarization observables, such as analyzing powers and spin correlation parameter, were obtained at 220 MeV \cite{brash} and 290 MeV \cite{rahav}.
 The analyzing powers and spin correlations  were also studied at IUCF at the energy of 197 MeV  \cite{miller}. 

The aim  of these experiments was
to study   the helium internal structure. The simple relations between the helium wave function and
differential cross section and polarization observables in the frame of the plane-wave-impulse-approximation (PWIA),
give an opportunity to extract useful information about  the ground state spin structure of  helium.   
In order to study the high-momentum components of the ${^3He}$, the elastic backward scattering of $p~~{^3He}$
was investigated  at RCNP(Osaka).
Here the differential cross section and  spin correlation parameter $C_{yy}$
were measured  at proton energies of
200, 300, and 400 MeV \cite{hatanaka}.

Several  years ago an experiment to study  $dd\to {^3He} n$, $dd\to {^3H} p$
reactions was carried out at RIKEN \cite {ourexp_T20}, \cite{ourexp_ay270}. The vector and tensor analyzing
powers were obtained in a wide angular range at three deuteron kinetic
energies: 140, 200, and 270 MeV. 
Previously  the differential cross sections of the reactions ${^2H}(d,n){^3He}$ and  ${^2H}(d,p){^3H}$ 
 were measured in a wide angular range for incident deuteron momenta between 1.1 GeV/c and
 2.5 GeV/c \cite{bizard}. The  $dd\to ^3He n$ reaction was considered in the one-nucleon-exchange (ONE) framework  
 in ref. \cite{ourprop, ouryf}. High sensitivity of some of the polarization observables was shown to
 the spin structure of  the ${^3He}$.
However, the data  obtained at RIKEN are in disagreement with ONE predictions.
Only a small angular range, around $0^0-15^0$ and $165^0-180^0$, is reasonably
described by ONE mechanism. This result stimulated further theoretical
investigations of this reaction. 

The four-nucleon problem is topical up to now
in spite of many efforts to solve it.
Significant progress in the studies of the  $dd\to {^3He} n (tp)$ reactions was achieved at
 low energies ($E_d < 5$ MeV)\cite{PRC77}, \cite{PRC76}, \cite{PRC81}. Here the reasonable description of the experimental
 data was obtained both for the differential cross sections and for the polarization observables.

 Practical integral equations for the four-body scattering were
developed by Grassberger and Sandhas \cite{ags4}. 
In  this formulation the original operator relations were reduced to
effective two-body equations in two steps by employing separable
expansions both for the two-body and for the three-body subamplitudes.
 After the partial wave decomposition we deal with one-dimensional equations. 

This approach was applied in ref.\cite{agsdd} to study $dd\to pt (n~{^3He})$
reactions and $p~{^3He}$ elastic scattering at the energies up to 51.5 MeV. 
Here the first-order $K$-matrix
 approximation was applied to solve effective two-body equations.
The obtained results
reasonably describe the shape of the differential cross sections but 
fail to reproduce the second maximum in the differential cross section of the
 $dd\to pt (n~{^3He})$.
 Inclusion of the
 principal value part of the propagators and use of  different potentials  
 did not result in significant improvement \cite{sofi}.
  Nevertheless, the
carried out investigations have shown that
 the agreement between the theoretical
predictions and data improves with increasing the energy when the second maximum is not so evident.

At higher energy the four-nucleon problem was considered in ref.\cite{dddd}, where
the deuteron-deuteron elastic scattering was studied at 231.8 MeV. The approximation 
based on the lowest order terms in the Neumann series expansion of 
the AGS- equations,  was used to describe the
differential cross section and vector and tensor analyzing powers. The  obtained results have
demonstrated the underestimation of the differential cross section while the curves for
the deuteron analyzing powers reproduce the behaviour of the data at forward angles.  

In the present paper the $dd\to {^3 He}n$ reaction is studied at the deuteron energies
between 200 MeV and 520 MeV. We start our investigation from AGS equations for the four-body case
\cite{ags4} and  then iterate them up to the first order terms over the nucleon-nucleon 
t-matrix. In such a way we include not only ONE mechanism into consideration
but also the next term. It corresponds to the case when nucleons from different
deuterons interact with each other  and then form a three-nucleon bounded state and a free nucleon.  
The parameterization  based on the modern phase-shift analysis data
is applied to describe NN interaction.
 The partial wave decomposition  is not used in this approach.
 It allows us to avoid the problem related with convergence which is important at
the considered energies. 

The paper is organized as follows. Section 2 gives the general
formalism. Here the expansion of  the AGS equations is presented
for the $dd\to {^3 He}n$ reaction. In this section the ${^3 He}$ wave function is discussed,
and   the
description of the nucleon-nucleon interaction is presented.
The details of calculations of the scattering amplitude terms
are also given. The obtained results  are
discussed in Sect.3. The conclusions are contained in Sect.4.

\section{General formalism}
\label{sec:1}

Here we consider the reaction where four initial nucleons are bounded
in pairs forming two deuterons, and three final nucleons are bounded to the helium or tritium and one nucleon is free.
In  other words, we have the reaction of the $(2)+(2)\to (3)+(1)$ type. 

We write the transition operator $U(z)$ for our reaction
as it was offered by  Grassberger and Sandhas  \cite{ags4}:
\begin{eqnarray}
\label{U}
U_{\beta\alpha}(z)=(1-\delta_{\beta\alpha})(z-H_\alpha)+\sum_{ik\nsubseteq\beta}
T_{ik}(z)G_0(z)U_{ik,\alpha}(z)+
\sum_{ik\nsubseteq\beta}V_\alpha\delta_{\alpha,ik}~,
\end{eqnarray}
where $\alpha$ and $\beta$ denote two-cluster partitions of the four-particles.
Here these labels are referred  to  the initial and final states, respectively: 
\begin{eqnarray}
\label{state}
&&1)\alpha =(ij)(kl);~~V_\alpha\equiv V_{(ij)(kl)}=V_{ij}+V_{kl};~~~\Phi_{(ij)(kl)}=|\vec k_{ij},\vec k_{kl}>
|\psi_{ij}>|\psi_{kl}>
\\
&&2)\beta=(ijk);~~~~~V_\beta\equiv V_{(ijk)}=V_{ij}+V_{jk}+V_{ik};~~~\Phi_{(ijk)}=|\vec k_{ijk},\vec k_l>|\psi_{ijk}>.
\nonumber
\end{eqnarray}
In accordance with the AGS-formalism
the channel Hamiltonian is defined as a sum of the  free particles Hamiltonian $H_0$  and the
interaction potential:
\begin{eqnarray}
H_{\alpha(\beta)}=H_0+V_{\alpha(\beta)}.
\end{eqnarray}

The eigenfunctions of the channel Hamiltonian $|\Phi_\alpha>$ characterize  possible initial and final configurations. These functions
are  products of plane waves and internal wave functions $|\psi_\alpha>$.

 The operator $T_{ij}(z)$ in Eq.(\ref U) is a two-body  transition operator which satisfies the Lippmann-
 Schwinger equation:
 \begin{eqnarray}
&&T_{ij}(z)=V_{ij}+V_{ij}G_{0}(z)T_{ij}~,
\end{eqnarray}
where $G_0$ is the resolvent of the four-nucleon kinetic energy operator
 $G_0(z)=(z-H_0)^{-1} $.

The operator $U_{ik,\alpha}$ in Eq.(\ref{U}) corresponds to the case when
 the initial state $\alpha$ is 
determined as in Eq.(\ref {state}) and the final state is a combination of  two bounded nucleons $(ik)$ and two free nucleons.
This transition operator can be also defined from Eq.(\ref {U}) if we put the final state $\beta=(ik)$.
The notation $ik\nsubseteq\beta$ means that pair $(ik)$ is not  either equal to one cluster of $\beta$ or  contained in it. 

We deal with four identical nucleons and two identical deuterons in the initial state.
It means that  symmetrized wave functions both for the initial and final states, should be built.
Following ref.\cite{GW} we have constructed a wave function for the initial state where four nucleons
form two bounded states: 
\begin{eqnarray}
|\psi(12)\psi(23)>_s=\frac{1}{4!}\sqrt{\frac{4!}{2!2!}}\sum_Q Q|\psi(12)>_s|\psi(23)>_s.
\end{eqnarray}
By  $Q$  we denote all possible permutations of two nucleons.
For four particles we have $4!$  permutations of this kind that is reflected in the first factor. 
The second coefficient is from  normalization of the symmetrized wave function. 
The wave functions of  deuterons $\psi(ij)_s$  are also  antisymmetrized:
\begin{eqnarray}
|\psi(ij)>_s=\frac{1}{\sqrt{2}}[|\psi(ij)>-|\psi(ji)>].
\end{eqnarray}
Three nucleons in the final state are bounded and one nucleon is free. The 
corresponding symmetrized wave function is as follows:
\begin{eqnarray}
|\psi(123)4>_s=\frac{\sqrt{4}}{4!}\sum_Q Q|\psi(123)>_s|4>.
\end{eqnarray}
Here  three-nucleon state $(ijk)_s$ is also presented by the antisymmetrized wave function
\begin{eqnarray}
|\psi(123)>_s=\frac{1}{\sqrt{6}}[|\psi(123)>-|\psi(213)>+|\psi(231)>-|\psi(321)>+|\psi(312)>-|\psi(132)>].
\end{eqnarray}
After straightforward  calculations the reaction amplitude can be written as: 
\begin{eqnarray}
\label{ampl}
<n{^3He}|U|dd>=\frac{1}{\sqrt{6}}[<4,\psi(123)_s|U|\psi(12)_s\psi(34)_s>-
<1,\psi(234)_s|U|\psi(12)_s\psi(34)_s>].
\end{eqnarray}
The same way it is necessary to find two matrix elements of the transition operator $U$. 
We start to consider  the first of them. This term corresponds to  the case of $\beta=(ijk)=(123)$, $\alpha=(12)(34)$.
From Eq.(\ref{U}) we get 
\begin{eqnarray}
U_{(123),(12)(34)}(z)&=&(z-H_0)-V_{12}-V_{34}+T_{14}(z)G_0(z)U_{(14),(12)(34)}(z)+
\nonumber\\
&&T_{24}(z)G_0(z)U_{(24),(12)(34)}(z)+T_{34}(z)G_0(z)U_{(34),(12)(34)}.
\end{eqnarray}
This relation contains transition operators for another reaction type.
In the final state two particles are bounded and the other two are free, while
the initial state is the same as before. In order to derive expressions for these operators,
 it is convenient to rewrite Eq.(\ref {U}) in the following form:   
\begin{eqnarray}
\label{U1}
U_{\beta\alpha}(z)=(1-\delta_{\beta\alpha})(z-H_\beta)+\sum_{mn\nsubseteq\alpha}
U_{\beta,mn}(z)G_0(z)T_{mn}(z)+
\sum_{mn\nsubseteq\alpha}V_\beta\delta_{\beta,mn}.
\end{eqnarray}
Then by putting $\beta=(ij)$  we obtain
\begin{eqnarray}
&&U_{(14),(12)(34)}(z)=(z-H_0)+U_{(14),(13)}(z)G_0(z)T_{13}(z)+U_{(14),(14)}(z)G_0(z)T_{14}(z)+
\nonumber\\
&&\hspace{3cm}
U_{(14),(23)}(z)G_0(z)T_{23}(z)+U_{(14),(24)}(z)G_0(z)T_{24}(z)
\nonumber\\
&&U_{(24),(12)(34)}(z)=(z-H_0)+U_{(24),(13)}(z)G_0(z)T_{13}(z)+U_{(24),(14)}(z)G_0(z)T_{14}(z)+
\\
&&\hspace{3cm}
U_{(24),(23)}(z)G_0(z)T_{23}(z)+U_{(24),(24)}(z)G_0(z)T_{24}(z)
\nonumber\\
&&U_{(34),(12)(34)}(z)=(z-H_0)-V_{34}+U_{(34),(13)}(z)G_0(z)T_{13}(z)+U_{(34),(14)}(z)G_0(z)T_{14}(z)+
\nonumber\\
&&\hspace{3cm}
U_{(34),(23)}(z)G_0(z)T_{23}(z)+U_{(34),(24)}(z)G_0(z)T_{24}(z).
\nonumber
\end{eqnarray}
Iterating these equations only up to the first order term over T-matrix, 
we get the following sequence for the $U_{(123),(12)(34)}$-operator:
\begin{eqnarray}
U_{(123),(12)(34)}\approx(z-H_{12})+T_{14}(z)+T_{24}(z).
\end{eqnarray}
Likewise we derive the expression for the other transition operator in Eq.(\ref{ampl}):
\begin{eqnarray}
U_{(234),(12)(34)}\approx(z-H_{34})+T_{13}(z)+T_{14}(z).
\end{eqnarray}

Since the initial and final states are antisymmetrized, the contributions of the
$T_{24}$ and $T_{14}$ matrix elements are equal to each other. In order to show it,
we use the properties of the permutation operator: $P_{12}P_{12}=1$, $P_{12}T_{24}P_{12}=T_{14}$.
\begin{eqnarray}
&&<4,\psi(123)_s|T_{24}|\psi(12)_s\psi(34)_s>=<4,\psi(123)_s|P_{12}P_{12}T_{24}P_{12}P_{12}|\psi(12)_s\psi(34)_s>=
\nonumber\\
&=&<4,\psi(213)_s|T_{12}|\psi(21)_s\psi(34)_s> =<4,\psi(123)_s|T_{12}|\psi(12)_s\psi(34)_s>
\end{eqnarray}
It  also concerns 
$T_{13}$ and $T_{14}$ matrix elements in the exchange contribution:
\begin{eqnarray}
<1,\psi(234)_s|T_{13}|\psi(12)_s\psi(34)_s>&=&<1,\psi(234)_s|T_{14}|\psi(12)_s\psi(34)_s>.
\end{eqnarray}
On the other hand, using the permutation operator $P_{14}$ we can get the following useful relation:
\begin{eqnarray}
<1,\psi(234)_s|T_{14}|\psi(12)_s\psi(34)_s>&=&- <4,\psi(123)_s|P_{14}T_{14}|\psi(12)_s\psi(34)_s>.
\end{eqnarray}

It gives us an opportunity to join all terms with NN $T$-matrix into one.
In such a way Eq.(\ref{ampl}) can be reduced to the following: 
\begin{eqnarray}
\label{uampl}
&&<n{^3He}|U|dd>=\frac{1}{\sqrt{6}}[<4,\psi(123)_s|z-H_{12}|\psi(12)_s\psi(34)_s>-
\\
&&<1,\psi(234)_s|z-H_{34}|\psi(12)_s\psi(34)_s>+
2<4,\psi(123)_s|T_{14}^{sym}|\psi(12)_s\psi(34)_s>],
\nonumber
\end{eqnarray}
where antisymmetrized NN T-matrix is defined as $T_{14}^{sym}=(1-P_{14})T_{14}$.

In order to simplify the statement below, we divide the latter expression via three terms corresponding
to these contributions into the reaction amplitude:
\begin{eqnarray}
\label{ampl}
<n{^3He}|U|dd>&=&\frac{1}{\sqrt{6}}\delta (2E_d-E_h-E_n)\{{\cal J}^{ONE}_{dir}+
{\cal J}^{ONE}_{exch}+2{\cal J}^{SS}\}.
\end{eqnarray}
 All the calculations have been performed
in the center-of-mass. The following definitions are introduced for the momenta and energies of the deuterons, helium, and 
neutron:
\begin{eqnarray}
&&\vec P_1=-\vec P_2\equiv \vec P_d, ~~\vec P_h=-\vec p_n
\\
&&E_d=\sqrt{M_d^2+P_d^2},~~E_h=\sqrt{M_h^2+P_h^2},~~ E_n=\sqrt{m_N^2+p_n^2}.
\nonumber
\end{eqnarray}
Two first terms in Eqs.(\ref{uampl}),(\ref{ampl}) correspond to the one-nucleon-exchange (ONE) mechanism of the reaction.
We call the first of them as "direct" and the second one  as "exchange".
Here one of the deuterons breaks in a neutron and proton. One of the nucleons becomes free, while the other
interacts with the remained deuteron forming helium or tritium. 
Schematically it can be presented by diagrams  in Figs.1a and 1b. The latter term corresponds to 
single scattering (SS)  when two nucleons from different deuterons interact in the final state (Fig, 1c).
\begin{figure}
\centering
  \includegraphics[height=10cm, width=14cm]{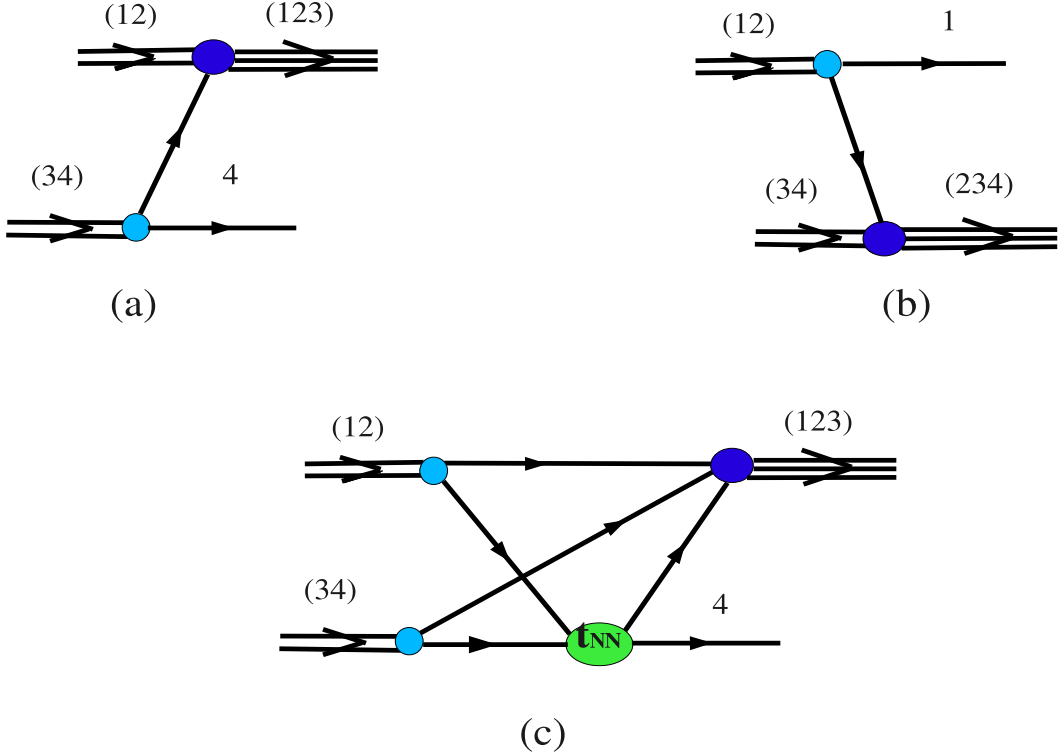}
\caption{The diagrams taken into consideration: one-nucleon-exchange (a),(b), and single
scattering (c) graphs.}
\label{fig:1}       
\end{figure}

\subsection{ One-nucleon-exchange}
\label{sec:2}

We start from consideration of  ONE contributions. Taking the quantum numbers and momenta of all particles
into account, we get the following expression for ONE terms:

\begin{eqnarray}
\label{one}
<n{^3He}|U|dd>&=&\frac{1}{\sqrt{6}}<\vec p_n m_n\tau_n|<\Psi^{123}(\vec P_h M_h\tau_h)|
(2E_d-\hat H_{12})|\Phi^{12}(\vec P_1,M_1)>|\Phi^{34}(\vec P_2,M_2)>-
\nonumber\\
&&<\vec p_n m_n\tau_n|<\Psi^{234}(\vec P_h M_h\tau_h)|
(2E_d-\hat H_{34})|\Phi^{12}(\vec P_1,M_1)>|\Phi^{34}(\vec P_2,M_2)>.
\end{eqnarray}

Here we introduce notation $\Psi^{ijk}(\vec P_h, M_h,\tau_h)$ for the $^3He$ wave function, where
$^3He$ is formed by $i,j,k$ nucleons and has momentum $\vec P_h$, spin projection $M_h$ and isospin projection 
$\tau_h$. Note in case $\tau=-1/2$ we deal with the reaction of $dd\to t p$. The $\Phi^{ij}(\vec P,M)$  denotes the 
wave function of the deuteron with momentum $\vec P$ and spin projection $M$.

Inserting the  unit operator into Eq.(\ref{one}):
\begin{eqnarray}
1=\int d\vec p_1 d\vec p_2 d\vec p_3 |\vec p_1 m_1\tau_1, \vec p_2 m_2\tau_2,\vec p_3 m_3\tau_3>
<\vec p_1 m_1\tau_1, \vec p_2 m_2\tau_2,\vec p_3 m_3\tau_3|~,
\end{eqnarray}
 for ONE contribution we get
\begin{eqnarray}
<n{^3He}|U|dd>&=&\frac{1}{\sqrt{6}}\int d\vec p_1 d\vec p_2 d\vec p_3 
\nonumber\\
&&(E_d-E_n-E_3)
<\Psi^{123}(\vec P_h M_h\tau_h)|\vec p_1 m_1\tau_1;\vec p_2 m_2\tau_2;\vec p_3 m_3\tau_3>
\nonumber\\
&&<\vec p_1 m_1\tau_1;\vec p_2 m_2\tau_2|\Phi^{12}(\vec P_d,M_1)><\vec p_3 m_3\tau_3;\vec p_n m_n\tau_n|\Phi^{34}(-\vec P_d,M_2)>-
\nonumber\\
&-&\frac{1}{\sqrt{6}}\int d\vec p_2 d\vec p_3 d\vec p_4
\\
&&(E_d-E_n-E_2)
<\Psi^{234}(\vec P_h M_h\tau_h)|\vec p_2 m_2\tau_2;\vec p_3 m_3\tau_3;\vec p_4 m_4\tau_4>
\nonumber\\
 &&<\vec p_n m_n\tau_n;\vec p_2 m_2\tau_2|\Phi^{12}(\vec P_d,M_1)><\vec p_3 m_3\tau_3;\vec p_4 m_4\tau_4|\Phi^{34}(-\vec P_d,M_2)>~.
\nonumber
\end{eqnarray}

\begin{table}[t]
\caption{Quantum numbers of the partial waves included into the definition of the three-body wave function.
$T, S, {\cal L}, J$ refer to the isospin, spin, orbital momentum and total angular momentum of the NN-subsystem.
$l$ is the relative orbital momentum of the spectator and $K$ is the channel spin \cite{he3}.}
\centering
\label{tab:1}
\begin{tabular}{llcllllll}
\hline\noalign{\smallskip}
 $\nu$ & Label & Subsystem & ${\cal L}$ & S & $J^\pi$ & $T$ & K & l  \\[3pt]
\hline
 1 & $^1s_0S$ & $^1s_0$ & 0 & 0 & $0^+$ & 1 & 1/2 & 0 \\
 2 & $^3s_1S$ & $^3s_1$ & 0 & 1 & $1^+$ & 0 & 1/2 & 0 \\
 3 & $^3s_1D$ & $^3s_1$ & 0 & 1 & $1^+$ & 0 & 3/2 & 2 \\
 4 & $^3d_1S$ & $^3d_1$ & 2 & 1 & $1^+$ & 0 & 1/2 & 0 \\
 5 & $^3d_1D$ & $^3d_1$ & 2 & 1 & $1^+$ & 0 & 3/2 & 2 \\
\noalign{\smallskip}\hline
\end{tabular}
\end{table}

Henceforth, we imply summations over all dummy discrete indices. 

 In our calculation we use the parameterized  wave function of a three-nucleon system offered in ref.\cite{he3}.
 This wave function was derived by fitting the full Faddeev wave function obtained with the CD Bonn \cite{cd} and
 Paris \cite{paris} NN-potentials.
 The wave function is fully antisymmetrized
and defined in terms of the nucleon pair and spectator momenta. If we choose particles (12) as a pair
and particle 3 as a spectator, the three-nucleon wave function is presented in the following form:
\begin{eqnarray}
\label{he3}
<\vec p_0 \vec q \nu|\Psi>&=&<\vec p_0 \vec q \nu|\psi[(12)3]>+
<\vec p_0 \vec q \nu|\vec p_{23}\vec q_1 \nu_{23}>
<\vec p_{23}\vec q_1 \nu_{23}|\psi[(23)1]>+
\\
&+&<\vec p_0 \vec q \nu|\vec p_{31}\vec q_2 \nu_{31}>
<\vec p_{31}\vec q_2 \nu_{31}|\psi[(31)2]>.
\nonumber
\end{eqnarray}

Here the following notations have been introduced for pair relative momentum $\vec p_0$ and spectator
momentum $\vec q$
\begin{eqnarray}
\label{jacobi}
\vec p_0&=&\frac{\vec p_1-\vec p_2}{2},~~~~~\vec q=\vec p_3-\frac{\vec P}{3},
~~~~\vec P=\vec p_1+\vec p_2+\vec p_3=\vec P_h
\nonumber\\
\vec p_{23}&=&\frac{\vec p_2-\vec p_3}{2},~~~~~\vec q_1=\vec p_1-\frac{\vec P}{3}
\\
\vec p_{31}&=&\frac{\vec p_3-\vec p_1}{2},~~~~~\vec q_2=\vec p_2-\frac{\vec P}{3}~.
\nonumber
\end{eqnarray}

The radial part of the three-nucleon wave function (\ref{he3}) is presented as a sum of the two terms each of them  has  a separable form:
\begin{eqnarray}
< p_0 q \nu|\Psi>=v^\nu_1 (p_0)w^\nu_1 (q)+v^\nu_2 (p_0)w^\nu_2 (q),
\end{eqnarray}
where $v^\nu_\lambda (p_0)$, $w^\nu_\lambda$ are defined as: 
\begin{eqnarray}
v^\nu_\lambda (p_0)=\sum_{n=1}^{5}\frac{a^\nu_{n,\lambda}}{p_0^2+(m^\nu_{n,\lambda})^2}~,~~~~~
w^\nu_\lambda (q)=\sum_{n=1}^{5}\frac{b^\nu_{n,\lambda}}{q^2+(M^\nu_{n,\lambda})^2}~,~~~~~~~~~\lambda=1,2~.
\end{eqnarray}
Index $\nu$ denotes  number of one of the three-nucleon channels (Table 1). The five channels are included into the definition of the wave function: ${^1s}_0 S, {^3s}_1 S, {^3}s_1 D,{^3d}_1S, {^3d}_1 D$. Parameters
$a^\nu_{n,\lambda}, b^\nu_{n,\lambda}, m^\nu_{n,\lambda}$ and $M^\nu_{n,\lambda}$ can be found in \cite{he3}.

The wave function of the deuteron, which contains $(ij)$ nucleons, is denoted in Eq.(\ref{one}) as 
$\Phi (\vec P_d, M_d)$. Here $\vec P_d$ and $M_d$ are momentum  and  spin projection of the deuteron, respectively.
In the rest frame the non-relativistic wave function  of the deuteron 
depends only on one variable $\vec p_0$ which is the
 relative momentum of the   proton- neutron pair:
\begin{eqnarray}
\label{dwf}
&&<\vec P_d/2+\vec p_0 ,
m_p;\vec P_d/2- \vec p_0 m_n|\Phi(\vec P_d M_d)>=
\\
&&\sum_{L=0,2} u_L(p_0)<\frac{1}{2} m_p \frac{1}{2} m_n|1 M_s>
<L M_L 1 M_s|1 M_d> Y_L^{M_L}(\hat p_0)~,
\nonumber
\end{eqnarray}
where $u_0(p_0)$ and $u_2(p_0)$ describe  $S$ and $D$ components of 
the deuteron wave function  \cite{cd}, \cite{par}, $\hat p_0$ is 
the unit vector in $\vec p_0$ direction, and $m_p,~m_n$ are the proton and neutron spin projections,
respectively.

Using transformations of vectors $\vec p_1, ~\vec p_2,~ \vec p_3$ to Jacobi variables (\ref{jacobi}) and taking
into account  momentum
conservations in the deuterons and helium vertices, we get the following expression for the first term
in Eq.(\ref{one}):
\begin{eqnarray}
\label{dir}
{\cal J}^{ONE}_{dir}&=&<\frac{1}{2}M_h \frac{1}{2} m_n|T^{ONE}_{dir}|1 M_1 1M_2>=
\nonumber\\
&&\frac{(-1)^{1/2-\tau_h}}{\sqrt{2}}{\cal K}\int d\hat p_0 dp_0 p_0^2 \psi^\nu (p_0,q){Y_l^\mu}^* (\hat q) {Y_{\cal L}^{\cal M}}^*(\hat p_0)
<{\cal L M} 1 M_S|J M_J>
\\
&&<J M_J\frac{1}{2} m_3|K M_K>
<l \mu  K M_K|\frac{1}{2} M_h><\frac{1}{2} m_3\frac{1}{2} m_n|1 M^\prime_S>
<L M_L 1 M_S|1 M_1> 
\nonumber\\
&&<L^\prime M^\prime_L 1 M^\prime_S|1 M_2> u_L(p_0)  Y_L^{M_L}(\hat p_0) 
u_{L^\prime} (|\vec P_h-\vec P_d/2|) Y_{L^\prime}^{M_L^\prime}(\widehat {\vec P_h-\vec P_d/2})
\nonumber
\end{eqnarray}
with kinematical factor ${\cal K}$ defined as
\begin{eqnarray}
{\cal K}=E_d-E_n-\sqrt{m_N+(\vec P_h-\vec P_d)^2}.
\end{eqnarray}
Definitions Eq.(\ref{he3}) and (\ref{dwf}) have been also used to obtain this equation. 
Superscribe index $\nu $ of the helium wave function marks one of the five channels considered
in \cite{he3} and defined by quantum numbers of the nucleon pair $({\cal L}, J)$, the relative orbital momentum of the spectator $l$ and  the channel spin $K$ \cite{chan} (Table 1).
 We also preserve here the dependence on isotopic number $\tau_h$ that allows us to consider both
$dd\to {^3He} n$ and  $dd\to t p$ reactions.
 As it follows from Eq.(\ref{jacobi}), spectator momentum $\vec q$ 
 is defined only by helium and deuteron momenta, $\vec q=\frac{2}{3}\vec P_h-\vec P_d$.
Since only two spherical functions in Eq.(\ref {dir}) are dependent of integration angles, we can simply
integrate this expression over the angular dependence of $\vec p_0$:
\begin{eqnarray}
&&{\cal J}^{ONE}_{dir}=
\frac{(-1)^{1/2-\tau_h}}{\sqrt{2} }{\cal K}\int d p_0 p_0^2 \psi^\nu (p_0,q) u_L(p_0) u_{L^\prime} (|\vec P_h-\vec P_d/2|)
< 1 M_1\frac{1}{2} m_3|K M_K>
\\
&&<l \mu K M_K|\frac{1}{2} M_h>
<L^\prime M^\prime_L 1 M^\prime_S|1 M_2>
<\frac{1}{2} m_3\frac{1}{2} m_n|1 M^\prime_S>
Y_{L^\prime} ^{M^\prime_L}(\widehat {\vec P_h-\vec P_d/2}){Y_l^\mu}^* (\hat q)~.
\nonumber
\end{eqnarray}

After substitution of the partial wave decomposition of the helium wave function \cite{he3}, we get the final
expression for the direct term of the ONE-contribution:  

\begin{eqnarray}
&&{\cal J}^{ONE}_{dir}=\frac{(-1)^{1/2-\tau_h}}{\sqrt{2}}{\cal K}\int dp_0 p_0^2 \{\frac{u_0(|\vec P_h -\vec P_d/2|)}{\sqrt{4\pi}}
<\frac{1}{2} m_3\frac{1}{2} m_n|1 M_2>+ 
\\
&&u_2(|\vec P_h -\vec P_d/2|)Y_2^{M^\prime_L}(\widehat {\vec P_h-\vec P_d/2})
<2 M^\prime_L 1 M^\prime_S|1 M_2><\frac{1}{2} m_3\frac{1}{2} m_n|1 M^\prime_S>\}
\nonumber\\
&&\{\frac{1}{\sqrt{4\pi}}[u_0(p_0)\psi_2(p_0,q)+u_2(p_0)\psi_4(p_0,q)]<1 M_1 \frac{1}{2} m_3|\frac{1}{2} M_h>
+
\nonumber\\
&&[u_0(p_0)\psi_3(p_0,q)+u_2(p_0)\psi_5(p_0,q)]Y_2^{\mu ~*} (\hat q)<1 M_1 \frac{1}{2} m_3|\frac{3}{2} M_K>
< 2 \mu\frac{3}{2} M_K|\frac{1}{2} M_h>\}~.
\nonumber
\end{eqnarray}
This expression contains only four components of the helium wave function, since channel $\nu =1$ corresponds
 to the  isotriplet state of the pair which is forbidden for the ONE-mechanism. 

In order to get  the exchange term of the ONE-amplitude ${\cal J}^{ONE}_{exch}$, it is necessary to replace  
$\vec P_d\to -\vec P_d$ and $M_1\longleftrightarrow M_2$ in the previous expression.

\subsection{Single scattering}
\label{sec:3}

The single-scattering term in Eq.(\ref{uampl}) can be rewritten in a more evident form:
\begin{eqnarray}
{\cal J}^{SS}&=&\int d\vec p_1 d\vec p_2 d\vec p_3 d\vec p_4 d\vec p_1^\prime <\Psi^{123}(\vec P_h M_h\tau_h)|\vec p_1^\prime m_1^\prime\tau_1^\prime;\vec p_2 m_2\tau_2;\vec p_3 m_3\tau_3>
\nonumber\\
&&<\vec p_1^\prime m_1^\prime\tau_1^\prime;\vec p_n m_n\tau_n|T(2E_d-E_2-E_3)|
\vec p_1 m_1\tau_1;\vec p_4 m_4\tau_4>
\\
&&<\vec p_1 m_1\tau_1;\vec p_2 m_2\tau_2|\Phi^{12}(\vec P_d,M_1)><\vec p_3 m_3\tau_3;\vec p_4 m_4\tau_4|\Phi^{34}(-\vec P_d,M_2)>.
\nonumber
\end{eqnarray}
We have here five integration vectors but three of them can be removed due to  the momentum conservation. 
We introduce vectors $\vec k_0$ and $\vec k_0^\prime$ which correspond to the neutron-proton
relative momenta in the deuterons:
\begin{eqnarray}
\vec k_0=\frac{1}{2}(\vec p_1-\vec p_2), ~~~~\vec k_0^\prime=\frac{1}{2}(\vec p_3-\vec p_4).
\end{eqnarray}
As it is mentioned above, we have used the three-nucleon wave function in a separable form which depends
on two variables: $\vec p_0$, a relative momentum of a pair, and spectator momentum $\vec q$.
In our calculations it is convenient to choose the nucleon pair (23) as a cluster and  
 nucleon 1 as a spectator. It is possible since our wave function is symmetrized:$\Psi^{123}=\Psi^{231}$.
Then the arguments of the helium wave function are expressed via momenta $\vec k_0$ and 
$\vec k_0^\prime$:
\begin{eqnarray}
\vec p_0=\frac{1}{2}(\vec P_d-\vec k_0-\vec k_0^\prime)~~~~
\vec q=\frac{2}{3}\vec P_h+\vec k_0-\vec k_0^\prime.
\end{eqnarray}

Using the definitions of $^3He$ and deuteron wave functions, Eqs.(\ref {he3} ),(\ref {dwf}), we
can write the following expression for the SS-term:
\begin{eqnarray}
&&{\cal J}^{SS}=<\frac{1}{2}M_h \frac{1}{2} m_n|T^{SS}|1 M_1 1M_2>=
\nonumber\\
&&\frac{(-1)^{1-\tau_1-\tau_3}}{2}
\int d\vec k_0 d\vec k_0^\prime \psi^\nu (p_0,q)Y_l^\mu(\hat q)^* Y_{\cal L}^{\cal M_L}(\hat p_0)^*
 <l \mu K M_K|\frac{1}{2} M_h>
\nonumber\\
&&<{\cal L M_L} S M_S|J M_J><J M_J \frac{1}{2} m_1^\prime|K M_K><\frac{1}{2} m_2\frac{1}{2} m_3|S M_S>
\nonumber\\
&& <T M_T\frac{1}{2} \tau_1^\prime|\frac{1}{2}\tau_h>
<\frac{1}{2} -\tau_1\frac{1}{2} \tau_3|T M_T>
<\frac{1}{2} \tau_1^\prime \frac{1}{2} -\tau_h|T^\prime M^\prime_T> <\frac{1}{2} \tau_1 \frac{1}{2} -\tau_3|T^\prime M^\prime_T>
\nonumber\\
&&
<\vec P_h+\vec k_0-\vec k_0^\prime, m_1^\prime;-\vec P_h, m_n |T(2E_d-E_2-E_3)|\vec P_d/2+\vec k_0, m_1; -\vec P_d/2-\vec k_0^\prime, m_4>
\nonumber\\
&&<\frac{1}{2} m_1\frac{1}{2} m_2|1 {\cal M}><L M_L 1{\cal M}|1 M_1>u_L(k_0)Y_L^{M_L}(\hat k_0)
\\
&& <\frac{1}{2} m_3\frac{1}{2} m_4|1 {\cal M}^\prime>
<L^\prime M_L^\prime 1{\cal M}^\prime|1 M_2> 
u_{L^\prime}(k_0^\prime)Y_{L^\prime}^{M_L^\prime}(\hat k_0^\prime)~.
\nonumber
\end{eqnarray}

The nucleon-nucleon scattering is described by the T-matrix element. 
 We use the parameterization  of this
 matrix offered by Love and Franey  \cite {LF}. This is the on-shell NN T-matrix
  defined in the center-of-mass:
\begin{eqnarray}
\label{tnn}
&&<\mbox{\boldmath$\kappa$}^{*\prime}  \mu_1^\prime \mu_2^\prime |t_{c.m.}|
\mbox{\boldmath$\kappa$}^* \mu_1\mu_2>
=<\mbox{\boldmath$\kappa$}^{*\prime}  \mu_1^\prime \mu_2^\prime |
A+B(\mbox{\boldmath$\sigma_1$} \hat N^*)(\mbox{\boldmath$\sigma_2$} \hat N^*)+
\\
&&
C(\mbox{\boldmath$\sigma_1$} +\mbox{\boldmath$\sigma_2$} )\cdot \hat N^* +
D(\mbox{\boldmath$\sigma_1$} \hat q^*)(\mbox{\boldmath$\sigma_2$} \hat q^*) +
F(\mbox{\boldmath$\sigma_1$} \hat Q^*)(\mbox{\boldmath$\sigma_2$} \hat Q^*)
|\mbox{\boldmath$\kappa$}^* \mu_1\mu_2>.
\nonumber
\end{eqnarray}
The orthonormal basis $\{\hat q^*,\hat Q^*,\hat N^*\}$ is a combination of the nucleon
relative momenta in the initial \mbox{\boldmath$\kappa $}$^*$ and final 
\mbox{\boldmath$\kappa$}$^{\prime *}$
states:
\begin{equation}
\hat q^*=\frac {\mbox{\boldmath $\kappa$}^* -\mbox{\boldmath$\kappa$}^{*\prime}}
{|\mbox{\boldmath$\kappa$}^* -\mbox{\boldmath$\kappa$}^{*\prime}|},~~
\hat Q^*=\frac {\mbox{\boldmath$\kappa$}^* +\mbox{\boldmath$\kappa$}^{*\prime} }
{|\mbox{\boldmath$\kappa$}^* +\mbox{\boldmath$\kappa$}^{*\prime}|},~~
\hat N^*=\frac {\mbox{\boldmath$\kappa$}^*  \times 
\mbox{\boldmath$\kappa$}^{*\prime} }{|\mbox{\boldmath$\kappa$}^*
\times\mbox{\boldmath$\kappa$}^{*\prime} |}.
\end{equation}
The amplitudes $A,B,C,D,F$ are the functions of the center-of-mass
energy and scattering angle. The radial parts of these amplitudes are taken 
as a sum of Yukawa terms. A new fit of the model parameters \cite{newlf} was done  
in accordance with the  phase-shift-analysis data SP07 \cite{said}.

Since the matrix elements are expressed
via the effective $NN$-interaction operators sandwiched
between the initial and final plane-wave states, this construction 
can be extended to the off-shell case allowing the initial and final
states to get the current values of  \mbox{\boldmath $\kappa$} and
\mbox{\boldmath $\kappa^\prime$}. Obviously, this extrapolation does 
not change the general spin structure.

In order to relate c.m.s. and the frame of our calculations, first of all, 
we apply Lorentz transformations to  kinematical variables. Let us consider
momenta and energies of the colliding nucleons:
\begin{eqnarray}
\label{lorentz}
\vec p_1&=&\mbox{\boldmath $\kappa$}^*+\vec u\left (\frac{(\vec u\mbox{\boldmath $\kappa$}^* )}{\gamma +1}+E^*\right ),
~~~~~E_1=\gamma E^*+(\vec u\mbox{\boldmath $\kappa$}^* )
\\
\vec p_4&=&-\mbox{\boldmath $\kappa$}^*+\vec u(\frac{-(\vec u\mbox{\boldmath $\kappa$}^* )}{\gamma +1}+E^*),
~~~E_4=\gamma E^*-(\vec u\mbox{\boldmath $\kappa$}^* ),
\nonumber
 \end{eqnarray} 
where $E^*$ is the energy of one of the nucleons in c.m.s. 
By  $u=(\gamma,\vec u)$ we denote  the 4-velocity of the reference frame relatively  c.m.s:
\begin{eqnarray}
\vec u=\frac{\vec p_1+\vec p_4}{\sqrt s},~~\gamma=\frac{E_1+E_4}{\sqrt s}.
\end{eqnarray}
Mandelstam variable $s$ is defined as usual:
\begin{eqnarray}
\sqrt s=2E^*=\sqrt{(E_1+E_4)^2-(\vec k_0-\vec k_0^\prime)^2}
\end{eqnarray}
Then two-nucleon state in the reference frame can be related with that in  the c.m.s. due to rotations in the spin
space of these nucleons: 
\begin{eqnarray}
|\vec p_1 m_1>|\vec p_4 m_4>={\cal N}|\mbox{\boldmath$\kappa$}^* m_1^\prime> |-\mbox{\boldmath$\kappa$}^* m_4^\prime>
D_{m_1^\prime m_1 }(\vec u, \vec p_1) D_{m_4^\prime m_4}(\vec u, \vec p_4),
\end{eqnarray}
where the Wigner rotation operator is
\begin{eqnarray}
D(\vec u, \vec p)=exp\{-i\vec\sigma\vec\omega\}=cos\frac{\omega}{2}\left ( 1-i\frac{\sigma[\vec u\times \vec p]}{(1+\gamma) m +E+\sqrt s}\right ).
\end{eqnarray}
The rotation is performed   around the axis
$[\vec u\times \vec p]$ on the angle $\omega$ determined by
\begin{eqnarray}
tg\frac{\omega}{2}=\frac{|\vec u\times \vec p|}{(1+\gamma) m +E+\sqrt s}
\end{eqnarray}
After transformations (\ref{lorentz}) the two-nucleon relative momentum in c.m.s. $\mbox{\boldmath$\kappa$}^*$ is written as follows: 
\begin{eqnarray}
&&\mbox{\boldmath$\kappa$}^*=\frac{\vec P_d}{2}+\frac{\vec k_0(E_4+\sqrt s)+\vec k_0^\prime(E_1+\sqrt s/2 )}
{E_1+E_4+\sqrt s}~.
\end{eqnarray}
Likewise we can obtain an expression for the relative momentum of the scattered nucleon pair:   
\begin{eqnarray}
&&\mbox{\boldmath$\kappa$}^{*\prime}=\vec P_h+\frac{(\vec k_0-\vec k_0^\prime)(E_n+\sqrt {s^\prime}/2)}
{E_1^\prime+E_n+\sqrt {s^\prime}}~.
\end{eqnarray}
Here we use the following definitions:
\begin{eqnarray}
&&E_1=\sqrt{m^2+(\vec k_0+\vec P_d/2)^2},~~~~~~~E_4=\sqrt{m^2+(\vec k_0^\prime+\vec P_d/2)^2},
\\
&&E_1^\prime=\sqrt{m^2+(\vec P_h^2+\vec k_0-\vec k_0^\prime)^2},~~~~
\sqrt {s^\prime}=2 E^{\prime *}=\sqrt{(E_1^\prime+E_n)^2-(\vec k_0-\vec k_0^\prime)^2}
\nonumber
\end{eqnarray}

It is also necessary to take into account normalization factor ${\cal N}$ and a kinematical
coefficient due to the transformation of the off-energy-shell $T$-matrix \cite{garc}.
The expressions for these factors were given in detail in ref.\cite{fb}.

It should be noted, that the current parameterization  describes the NN-interaction in the wide energy range
between 50 MeV and 1100 MeV \cite{newlf}. 
However, at low energies ($<100$ MeV) the quality of the parameterization can not be assessed due to the lack of the experimental data. Therefore, we do not
consider the single scattering contribution into  $dd\to {^3}He n$ reaction amplitude at the deuteron energy below 200 MeV, where the Faddeev calculation technique is
more preferable.

\begin{figure}
\centering
  \includegraphics[height=10cm, width=14cm]{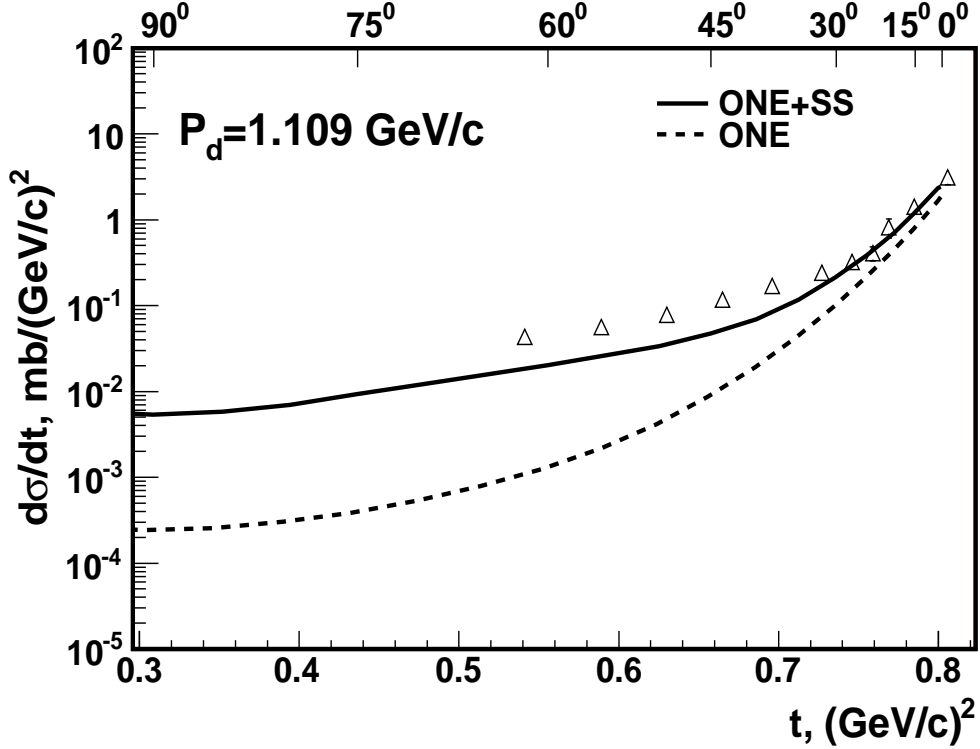}
\caption{ The differential cross section at the deuteron momentum  
 of 1.109 GeV/c as a function of t.
The data are taken from \cite{bizard}.}
\label{fig:2}       
\end{figure}

\begin{figure}
\centering
  \includegraphics[height=10cm, width=14cm]{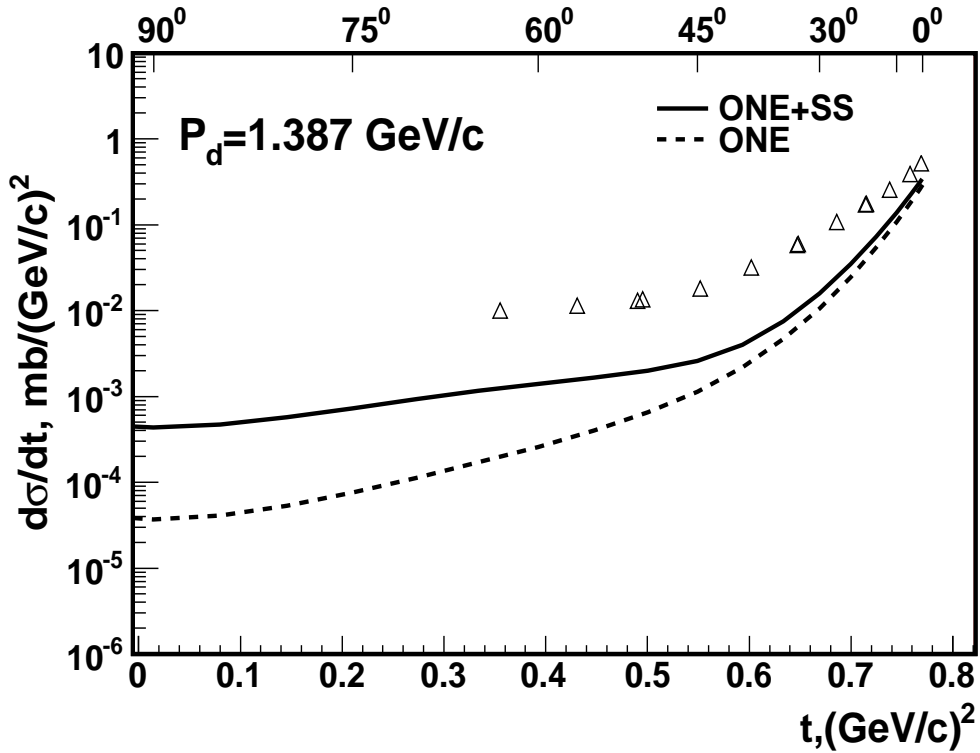}
\caption{ The differential cross section at the deuteron momentum  
 of 1.387 GeV/c as a function of t.
The data are taken from \cite{bizard}.}
\label{fig:3}       
\end{figure}

\section{ Results and discussions}
\label{sec:5}

The formalism presented above was applied to describe the experimental data 
obtained for $dd\to {^3 He n}$ and $dd\to t p$ reactions at the deuteron kinetic energies
of a few hundred MeV. The calculations have been performed with CD-Bonn deuteron and helium wave functions. 
 The differential cross section can be written as a function of 
Mandelstam variables $s$ and $t=(P_d-P_{He})^2$:

\begin{eqnarray}
\frac{d\sigma}{dt}=\frac{4\pi}{9 s}\frac{s^2-(M^2_{He}-m^2)^2}{s-4M_d^2}|{\cal J}(s,t)|^2~.
\end{eqnarray}

 We consider three energies, 300 MeV, 457 MeV and 520 MeV, 
which correspond to the laboratory momenta $P_{lab}= 1.109,~ 1.387$, and $1.493$ GeV/c, respectively.
In this energy range the presented formalism is more successful. Moreover, we have a set of the experimental
data on the differential cross sections in a wide angular range obtained at these energies in Saclay \cite{bizard}.

 In Figs.2-4 the results of the calculations of the differential cross sections are presented in comparison
with the data.
In order to demonstrate the contribution of the single scattering term, we have considered two cases. One of them corresponds to the
calculations including only ONE terms. The results of these calculations are given with the dashed curves.
The other case corresponds to the calculations taking into account both ONE and single scattering contributions. 
These results are presented with the solid curves.

As expected, the contribution of the rescattering term is not  large at small scattering angles ($t\sim 0.7-0.8 (GeV/c)^2$).
It is in agreement with the results obtained in ref.\cite{ourexp_ay270}.
However, the difference between these two curves increases with the angle and reaches the maximal
value at $90^0$. Taking the single scattering diagram into consideration significantly
improves the agreement between the experimental data and theoretical predictions.
We have a good description of the data  for 
$P_{lab}=1.109 ~GeV/c$ (Fig.2).
 Nevertheless, the underestimation of the differential cross sections is observed 
 at the deuteron energies above 300 MeV (Figs.3,4).
 Perhaps, this discrepancy can be reduced, if the $\Delta$-excitation in the intermediate state
  is taken into account. This possibility is discussed in ref.\cite{bizard}, where
  the $\Delta$-isobar is taken into consideration in the simplest phenomenological model. 
  
The formalism  presented here gives us an opportunity to calculate not only the differential cross sections but also
polarization observables. In this paper we have considered the energy dependence of tensor analyzing power  
$T_{20}$ at the scattering angle
equal to zero (Fig.4).
 The experimental data were obtained at RIKEN \cite{ourexp_T20}. As it is mentioned above,
the contribution of the single scattering term is not large at small angles.  
 Nevertheless, one can observe
some improvement of the agreement between the data and theory predictions. Unfortunately, we do not have  enough
experimental data to
confirm this tendency.

\begin{figure}
\centering
  \includegraphics[height=10cm, width=14cm]{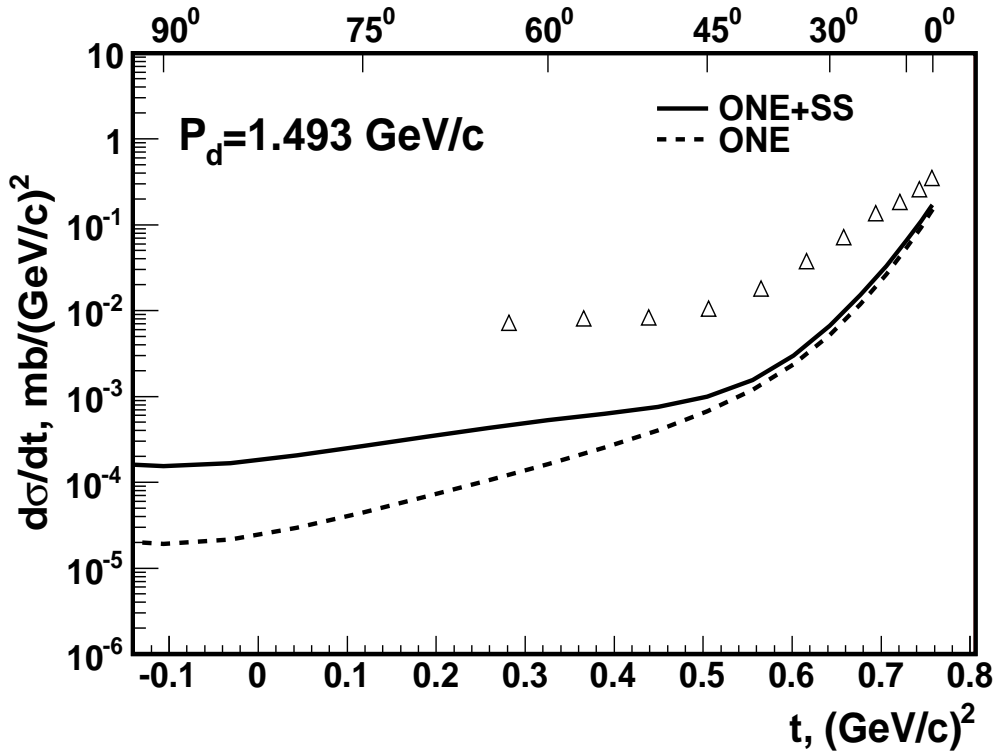}
\caption{ The differential cross section at the deuteron momentum  
 of 1.493 GeV/c as a function of t.
The data are taken from \cite{bizard}.}
\label{fig:4}       
\end{figure}

\begin{figure}
\centering
  \includegraphics[height=10cm, width=14cm]{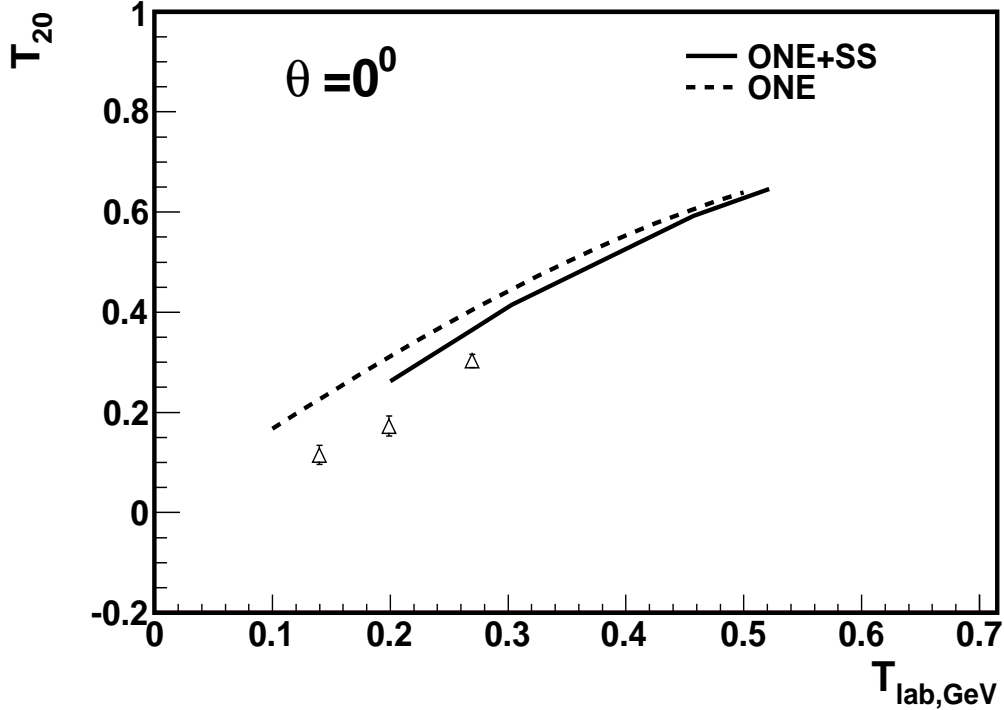}
\caption{The energy dependence of tensor analyzing power $T_{20}$ at the zero scattering angle.
The data are taken from \cite{ourexp_T20}}
\label{fig:5}       
\end{figure}

\section{Conclusions.}
\label{sec:6}

The model to describe  the $dd\to {^3He}~n(tp)$ reaction at the energies of a few
hundred MeV has been presented in this paper. We start from the AGS-equations for 
N-body system. 
Iterating these equations over the NN $t$-matrix we obtain the expression for the reaction amplitude.
In the presented calculations only two lowest terms of this expansion are included into this consideration.  
Here we do not solve any equations to define wave functions of the bounded states or to find the
nucleon-nucleon $t$-matrix. Instead of that the  parameterized wave functions for 
the deuteron and helium are used. These parameterizations 
 take the spin structure of these  nuclei into account. In order to describe 
  interactions of the nucleons in the intermediate state, 
the parameterized  $NN$ $t$-matrix is applied that allows us to avoid
the problem of  convergence which appears at the partial wave decomposition at these energies.

The presented model has been applied to describe differential cross
sections at deuteron energies of 300 MeV, 493 MeV, and 520 MeV. A reasonable agreement
between the data and theoretical results has been obtained for the energy equal to 300 MeV.
It is shown that the contribution of the single scattering term is small at the
forward scattering angles while inclusion of  the rescattering diagram significantly
improves the description of the experimental data at the scattering angles larger than $30^0$.
 The energy
dependence of the  $T_{20}$ has been also obtained at the energy range between 200 MeV and
520 MeV at the zero scattering angle. Some improvement of the data description has been received
when the single scattering term is taken into account.
All these results allow us to regard this approach as the next step in addition 
to the one-nucleon-exchange mechanism to
 solve the four-nucleon problem.

{\bf Acknowledgements}
The author is grateful to Dr. V.P. Ladygin for fruitful discussions.
This work has been supported by the Russian Foundation for Basic Research
under grant  $N^{\underline 0}$  10-02-00087a.



\end{document}